\documentclass{aa}

\usepackage{txfonts,upgreek}

\usepackage{natbib}
\usepackage{subfig}

\usepackage{graphicx}   

\usepackage{booktabs}
\usepackage{subfig}
\usepackage{hyperref}
\hypersetup{
  colorlinks,
  citecolor=blue,
  linkcolor=red,
  urlcolor=blue}

\bibpunct{(}{)}{;}{a}{}{,} 

\newcommand{\PSRCHIVE}{\texttt{PSRCHIVE}}
\newcommand{\TEMPO}{\texttt{TEMPO}}

\newcommand{\msun}{M$_{\sun}$}

\newcommand{\chisquare}{\chi^2_{\textrm{red}}}
\newcommand{\Pbdot}{\dot{P}_{\rm b}}

\newcommand{\Pdotobs}{\dot{P}_{\rm obs}}
\newcommand{\Pdotint}{\dot{P}_{\rm int}}
\newcommand{\Pbdotobs}{\dot{P}_{\rm b \rm ,obs}}

\newcommand{\Mp}{M_{\rm p}}
\newcommand{\Mc}{M_{\rm c}}
\newcommand{\Mtot}{M_{\rm tot}}

\begin{document}

\title{Detection of the relativistic Shapiro delay in a highly inclined millisecond pulsar binary PSR~J1012$-$4235}
\titlerunning{Shapiro delay in PSR J1012$-$4235}
\author{
T. Gautam \inst{1}\inst{2}\and
P. C. C. Freire\inst{1}\and
J. Wu\inst{1}\and
V. Venkatraman Krishnan\inst{1}\and
M. Kramer\inst{1}\and
E. D. Barr\inst{1}\and
M. Bailes\inst{3}\inst{4} \and
A. D. Cameron\inst{3}\inst{4}}
\institute{Max-Planck-Institut f\"{u}r Radioastronomie, Auf dem H\"{u}gel 69, D-53121 Bonn, Germany\label{1}\and
National Radio Astronomy Observatory, 520 Edgemont Rd., Charlottesville, VA, 22903, USA\label{2}\and
Centre for Astrophysics and Supercomputing, Swinburne University of Technology, PO Box 218, VIC 3122, Australia\label{3}\and
ARC Centre of Excellence for Gravitational Wave Discovery (OzGrav), Swinburne University of Technology, PO Box 218, VIC 3122, Australia \label{4}\\
\email{tgautam@nrao.edu}}

\date{}


\abstract
{PSR~J1012$-$4235 is a 3.1ms pulsar in a wide binary (37.9 days) with a white dwarf companion. We detect, for the first time, a strong relativistic Shapiro delay signature in PSR~J1012$-$4235. Our detection is the result of a timing analysis of data spanning 13 years and collected with the Green Bank, Parkes, and MeerKAT Radio Telescopes and the \textit{Fermi} $\gamma$-ray space telescope. We measured the orthometric parameters for Shapiro delay and obtained a 22$\sigma$ detection of the $h_{\rm 3}$ parameter of 1.222(54) $\upmu$s and a 200$\sigma$ detection of $\varsigma$ of 0.9646(49). With the assumption of general relativity, these measurements constrain the pulsar mass ($\Mp=1.44^{+0.13}_{-0.12}$ \msun), the mass of the white dwarf companion ($\Mc = 0.270^{+0.016}_{-0.015}$ \msun), and the orbital inclination ($i=88.06^{+0.28}_{-0.25} \deg$). Including the early $\gamma$-ray data in our timing analysis facilitated a precise measurement of the proper motion of the system of 6.58(5) mas yr$^{-1}$. We also show that the system has unusually small kinematic corrections to the measurement of the orbital period derivative, and therefore has the potential to yield stringent constraints on the variation of the gravitational constant in the future. }



\maketitle
\section{Introduction}
Due to their rotational stability and high spin frequencies, millisecond pulsars (MSPs) provide remarkable timing precision and have therefore proven to be excellent probes of gravity theories and the physics of dense matter in neutron stars (NSs; see \citealt{2016ARA&A..54..401O}), and have been used to constrain the $\rm nHz$ gravitational wave background \citep{2023ApJ...951L...8A,2023ApJ...951L...6R,2023A&A...678A..48E,2023RAA....23g5024X}. These diverse applications are a result of modelling the physical phenomena that influence the times of arrival (ToAs) of pulses at the telescope or receiver.

In the case of binary pulsars, the modelling of the ToAs can lead to the detection of several relativistic effects, which either appear as deviations from the Keplerian orbital motion of the pulsar, or as additions to the propagation time of the radio waves to the Earth. These relativistic effects can be quantified (and therefore parameterised) in a theory-independent way with the so-called `post-Keplerian (PK) parameters' \citep{PhysRevD.45.1840}. 
Assuming a theory of gravity, such as general relativity (GR), these PK parameters become functions of the Keplerian parameters of the  orbit of the pulsar as well as its mass and that of its companion. Therefore, measurements of two PK parameters can ---with the assumption of a specific gravity theory--- directly constrain the masses of the two component stars of the system. A detection of three or more PK parameters will provide additional ways to determine the same two masses, providing consistency tests of the gravity theory being assumed
(see e.g. \citealt{2021PhRvX..11d1050K} and references therein). 

In this work, we discuss our timing analysis of the MSP system PSR~J1012$-$4235, and our detection of a Shapiro delay \citep{1964PhRvL..13..789S} in its timing. This effect is observed when the radio pulses are delayed as they propagate through a space-time curved by the pulsar's companion; it is more easily observed in edge-on binaries, where the companion passes close to the pulsar's line of sight. The variation of the Shapiro delay with orbital phase allows the measurement of two PK parameters (see details below), which are sufficient for a determination of the component masses \citep{2023MNRAS.520.1789S}.

PSR~J1012$-$4235 is a 3.1ms pulsar discovered using the Parkes radio telescope in a survey \citep{2015ApJ...810...85C} that targeted unidentified $\gamma$-ray sources found with the Large Area Telescope (LAT; \citealt{2009ApJ...697.1071A}) on board the Fermi $\gamma$-ray space telescope. In the discovery paper, by fitting for the changes in the barycentric period due to changes in the Doppler shift (caused by the changing orbital velocity), the orbital parameters of the binary were determined, although with low precision. The pulsar has a low-eccentricity ($e < 0.001$) orbit with a period of 37.9 days around a companion with a minimum mass of $0.26$ \msun. This suggests it is likely a helium-core white dwarf (He WD). This minimum companion mass is very close to the companion mass predicted by \cite{1999A&A...350..928T} for a He WD companion for a system with this orbital period. This indicates that, if the companion is indeed a He WD and the model is correct, the system is probably being viewed edge-on, that is, with an orbital inclination ($i$) of close to $90\deg$.

From 2013 to 2015, additional timing observations were carried out with the 64m Parkes `Murriyang' radio telescope and the Green Bank Telescope (GBT), and a phase-connected timing solution was derived with much improved constraints on the position, spin-down, and orbital parameters.
The eccentricity and orbital period of the system were found to be consistent with the MSP-He WD relationship given by \cite{1994ARA&A..32..591P}. 


The detection of the Shapiro delay in this system, made possible by
the high sensitivity of the  MeerKAT telescope \citep{2016mks..confE...1J},
confirmed the high predicted orbital inclination. This 64-dish array provides excellent timing sensitivity for pulsars in the southern hemisphere \citep{2020PASA...37...28B}. With the aim of performing high-precision timing for a large number of pulsar systems, a Large Science Project (LSP) called MeerTIME \citep{2016mks..confE..11B,2020PASA...37...28B} is being carried out. One program under this project is the relativistic binary program (referred to as `RelBin'). RelBin specifically targets relativistic binary pulsars for the purpose of measuring relativistic effects, constraining NS masses, and testing GR, along with constraining alternative theories of gravity \citep{2021MNRAS.504.2094K}. Another MeerTIME program is the `pulsar timing array' (PTA), the aim of which is to use MSPs for the detection of nHz gravitational waves \citep{2021MNRAS.502..407P,2022PASA...39...27S,2023MNRAS.519.3976M}. The PSR~J1012$-$4235 system was observed regularly from 2019 to 2022 under the PTA and RelBin programs.



In this paper, we present a phase-connected timing solution for PSR J1012$-$4235, which in this case is derived from the timing observations obtained with the Parkes, GBT, and MeerKAT radio telescopes, and $\gamma$-ray data collected by Fermi-LAT. We obtained precise measurements of the proper motion, position, and orbital parameters.
The detection of a strong Shapiro delay signature was made possible by dedicated orbital campaigns carried out under the RelBin program, especially around superior conjunction and the extrema of the unabsorbed component of the Shapiro delay signal. This provides good measurements of the orbital inclination of the binary and the pulsar and companion masses.

The present paper is organised as follows: in Section~\ref{sec:observations_1012} we describe the radio data set used for the analysis in this work, in Section~\ref{sec:data_reduction} we discuss the data reduction procedure of both radio and $\gamma-$ray data, and in Section~\ref{sec:polcal} we discuss the polarisation profile of the pulsar and the fit of its linear polarisation using the Rotating Vector Model (RVM, \citealt{1969ApL.....3..225R}). In Section~\ref{sec:timing_analysis} we discuss the timing analysis and its results, in Section~\ref{sec:gammaandradio} we discuss the $\gamma-$ray pulse profile and compare its features with the radio profile, and finally in Section~\ref{sec:conclusion_1012} we summarise our results.

\section{Observations} 
\label{sec:observations_1012}

We now describe the radio data set analysed in this work.
The details of all the radio observations used in this work are listed in Table \ref{table:observations_1012_details}.

\subsection{Parkes and GBT observations}

Discovery observations and follow-up timing observations (continued until January 2015) of this pulsar were carried out with the Parkes radio telescope. Stokes-I data were recorded in search mode with the central beam of the 20 cm multi-beam receiver \citep{1996PASA...13..243S}. For a detailed description of this data set, we refer the reader to \cite{Kerr_2012} and \cite{2015ApJ...810...85C}.
\\
To further refine the timing solution of the pulsar, timing observations were also carried out with the GBT in 2013 and 2014. For these observations, search-mode data were obtained with the 820MHz receiver covering 200 MHz of bandwidth. Only Stokes-I data were recorded, which were sampled at 40.96 $\upmu$s with 2048 frequency channels.
\subsection{MeerKAT observations}
\begin{figure}
\centering
        \includegraphics[width=0.8\linewidth]{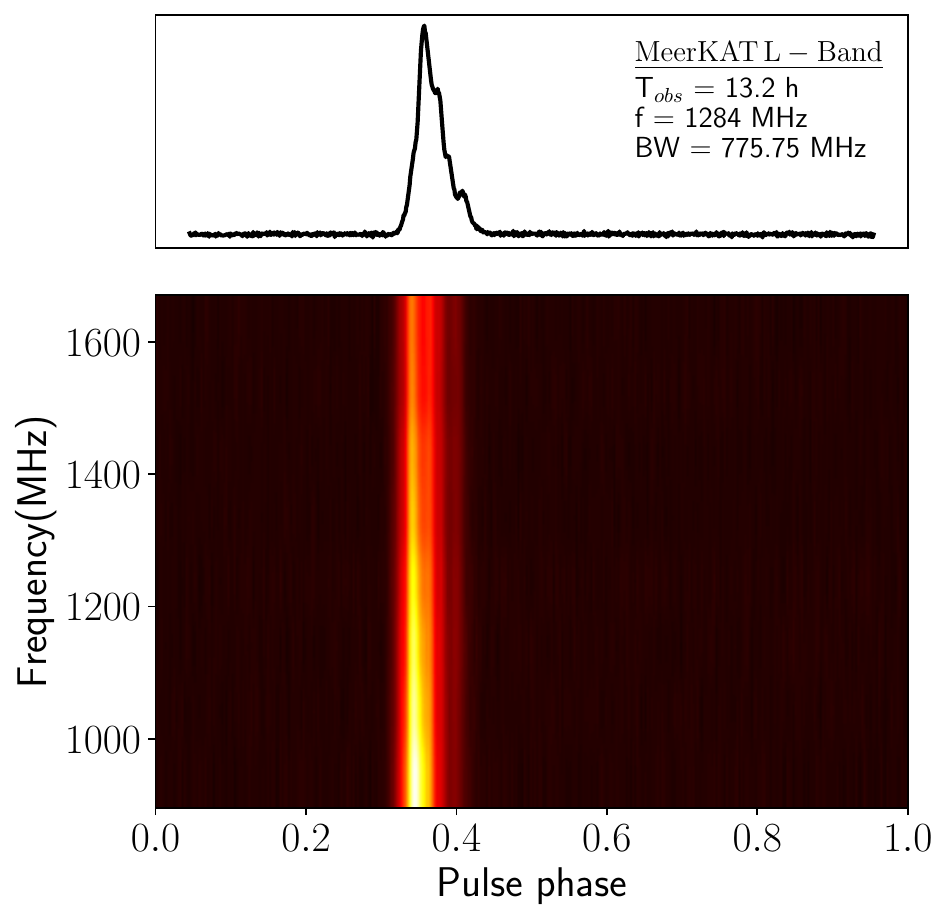}
        \caption{MeerKAT L-Band profile obtained by integrating a total of 13.2 hr of data. {\em Top plot}: Intensity versus rotation phase. {\em Bottom plot}: Radio frequency versus rotation phase. In this latter plot, we can see how the pulse profile evolves with radio frequency. The number of phase bins was 1024 and the S/N is 977.}
        \label{fig:mkt-profile}
\end{figure}

The instrumentation and setup for pulsar observations with MeerKAT are explained in detail in \cite{2020PASA...37...28B}. The timing observations with MeerKAT for this pulsar were performed as part of both PTA  and RelBin programs. All the observations were carried out with the L-band receiver and used the PTUSE backend \citep{2020PASA...37...28B}. Typical observation times of PTA observations were $\sim$ 5 min, while RelBin observations (which focused on getting a better orbital coverage) were typically $\sim$ 34 mins. In addition, we carried out one long observation of $\sim$ 5 hr covering superior conjunction under the RelBin program. All the observations covered an effective bandwidth of 775.75 MHz divided across 928 frequency channels. The MeerKAT data used in this work were taken from April 2019-January 2022, and amounted to a total of 73 observations covering 13.2 hr. The integrated pulse profile from MeerKAT observations is shown in Fig.~\ref{fig:mkt-profile}.  

\begin{table*}
\caption[]{Details of radio observations.}
\label{table:observations_1012_details}
\footnotesize
\centering
\renewcommand{\arraystretch}{0.5}
\setlength{\tabcolsep}{3.0pt}
\vskip 0.1cm
\begin{tabular}{lcccccccc}
\hline
{Observatory} & {Epoch} & {Receiver} & {Central frequency} & {Bandwidth} & Number of observations  & Total integration time & {\emph{\#}ToAs} & {EFAC} \\ {}& {(MJD)} & & {(MHz)} & {(MHz)} & & (hrs) & \\ \midrule \midrule
{Parkes}& {56469-57039} & 20-cm multibeam &  1382 & 400  & 22 & 9.9 & 169 & 1.22 \\
{}&   &  &   &  \\
{GBT}&{56522-56732} & 820-MHz receiver & 820 & 200  & 14 & 3.6 & 93 & 0.96 \\
{}&   &  &   &  \\
{MeerKAT}& 58595-59604 & L-band & 1283.582 & 775.75 & 73 & 13.2 & 3446 & 1.03 \\
\hline
\hline
\end{tabular}
\end{table*}




\section{Data reduction}
\label{sec:data_reduction}
\subsection{Radio data}
All the earlier Parkes and GBT data available in the folded \texttt{psrchive} format were reloaded with a consistent phase-connected timing solution in order to improve the signal-to-noise ratio (S/N) of the pulse profiles. To enhance the S/N of individual ToAs, each Parkes epoch was integrated in frequency, while two frequency sub-bands were formed for each GBT epoch.

The entire MeerKAT data reduction was performed using the standard pulsar analysis software \textsc{PSRCHIVE}\footnote{\url{https://psrchive.sourceforge.net/}}\citep{2004PASA...21..302H,2012AR&T....9..237V}. All the MeerKAT L-Band data were passed through a data-reduction pipeline: \textsc{MEERPIPE}\footnote{\url{https://github.com/OZGrav/meerpipe/}}\citep{2021MNRAS.502..407P}, which first performs the RFI excision using a modified version of \textsc{COASTGUARD} \citep{2016MNRAS.458..868L}, and then carries out the polarisation and flux calibration. In this process, to remove the low power levels at the edges of the bandpass, all the observations were reduced from their original bandwidth of 856 MHz to 775.75 MHz. The pipeline outputs a consistent data set, with similar bandwidth and central frequency, which can be readily used for the timing analysis. 

To calculate the pulse ToAs of all high-S/N integrations in the MeerKAT data, we cross-correlated them with a standard template of the pulse profile. As the intrinsic profile shape of a pulse can vary with frequency ---and this is certainly the case for PSR~J1012$-$4235, as we see in Fig.~\ref{fig:mkt-profile}---, using a single template created by integrating in frequency can smear some of the profile features and lead to increased ToA uncertainty. Therefore, a template with multiple frequency sub-bands can provide smaller ToA uncertainties. Additionally, sharp features in the profiles with different polarisation can help improve the timing precision further, and therefore a template with non-integrated polarisation can be useful in some cases. In order to identify the best-suited standard template for our case, and one that would give the smallest residual RMS from the fit, we created four different types of template and produced ToAs with each of them:

\begin{enumerate}
\item 1F1P: Simple one-dimensional profile, where we integrated in both frequency and polarisation, created with the \texttt{paas} routine of \PSRCHIVE.
\item 8F1P: Eight-frequency band profile, sub-banded to eight frequency channels and integrated completely in polarisation; to generate ToAs in this case we used the \texttt{Fourier domain with Markov chain Monte Carlo (FDM)} algorithm of the \texttt{pat} routine.
\item 1F4P: One-dimensional profile with polarization information. We integrated in frequency but with full Stokes parameter profiles created using the \texttt{matrix template matching (MTM)} algorithm of the \texttt{pat} routine to generate ToAs.
\item 8F4P: Sub-banded to eight frequency channels and non-integrated polarisation profiles with full Stokes parameters. 
\end{enumerate}
In addition to these, templates with four rather than eight frequency channels in schemes (ii) and (iv) were also created for comparison. 
We find that the best template for MeerKAT data that provides minimum parameter uncertainties for our dataset is (2), that is, integrated in polarisation but sub-banded to eight frequency channels. For the earlier dataset (both Parkes and GBT), the best resulting reduced $\chi^{2}$$(\chisquare)$ comes from individual templates formed with integrated frequency. 

Datasets taken from different telescopes (GBT, Parkes, and MeerKAT in this case) lead to additional time delays due to different instruments, which differ both in terms of their design and their geographic location. To account for these delays, we fit for an arbitrary phase offset (`JUMPS') between each telescope datum. Additionally, an inconsistent definition of templates can lead to unreliable prediction of orbital parameters and high timing parameter uncertainties (as discussed in detail in \citealt{2021A&A...654A..16G}). For this reason, we aligned all three standard templates created for the three datasets with a consistent phase reference point.

Typically, for MeerKAT data, we generated ToAs for one-minute subintegrations, leading to eight ToAs per minute. The total number of ToAs extracted from each data set is given in Table~\ref{table:observations_1012_details}. The Table also provides values for the EFAC parameter, which is used to rescale the uncertainties of ToAs in order to avoid unmodelled systematic effects; it was calculated so that the reduced $\chi^2$ of the fit for each data set is 1. This increase in the ToA uncertainties results in more conservative estimates of the uncertainties of the timing parameters.


\subsection{\texorpdfstring{$\gamma-$} .ray data}
\label{sec:gamma-data}
The Parkes survey by \cite{2015ApJ...810...85C} was designed to search for unidentified $\gamma-$ray sources. PSR~J1012$-$4235 was found in the $\gamma-$ray source 3FGL J1012$-$4235 presented in the third Fermi-LAT catalogue \citep{2015ApJS..218...23A}. 
With the preliminary timing solution obtained using only radio data, we can
see that, as is the case for $\sim$300 other pulsars \citep{2023arXiv230711132S},
PSR~J1012$-$4235 is pulsating in $\gamma$-rays, confirming the association between the radio pulsar and the $\gamma$-ray source. 

To determine the likelihood of each photon being a signal or background, we applied a simplified weighted method developed by \cite{bruel2019extending}, which considers factors such as photon energy and angular separation from the pulsar position. This method allowed us to assign a probability (or weight) to each photon for the length of the entire {\it Fermi} mission
---more specifically, from Mission Elapsed Time (MET) 239558000 (MJD 54682.66202546) and MET 667972000 (MJD 59641.15734954)--- within the 0.1-3 GeV energy range and within 1 degree of the pulsar position.

Using the weighted photon set, we calculated the pulse phase for every $\gamma-$ray photon based on the best radio timing solution available, using the TEMPO2 Fermi plugin\footnote{\url{https://fermi.gsfc.nasa.gov/ssc/data/analysis/user/}}. We examined the resulting folded $\gamma-$ray profile to assess the adequacy of the initial timing solution for further analysis.

As discussed in Section~\ref{sec:gammaandradio}, this folded $\gamma$-ray profile is characterised by a double peak consisting of strong and narrow pulses. These provide precise, albeit sparse $\gamma$-ray ToAs:
in total, we extracted 32 $\gamma-$ray ToAs (each with uniform S/N) using a multi-component Gaussian template employing the maximum likelihood methods described in \cite{ray2011precise}. These ToAs nicely complement the radio data, and are especially useful for constraining parameters with long-term trends in residuals, such as the spin-period derivative and the proper motion. Their addition especially helps to constrain long-term DM variations to which the $\gamma-$ray ToAs are immune. They also confirm that we have not missed a single rotation in the four-year gap of radio observations between 2015 and 2019.
 
\section{Polarisation}
\label{sec:polcal}

\begin{figure}
\centering
        \includegraphics[width=1.1\columnwidth]{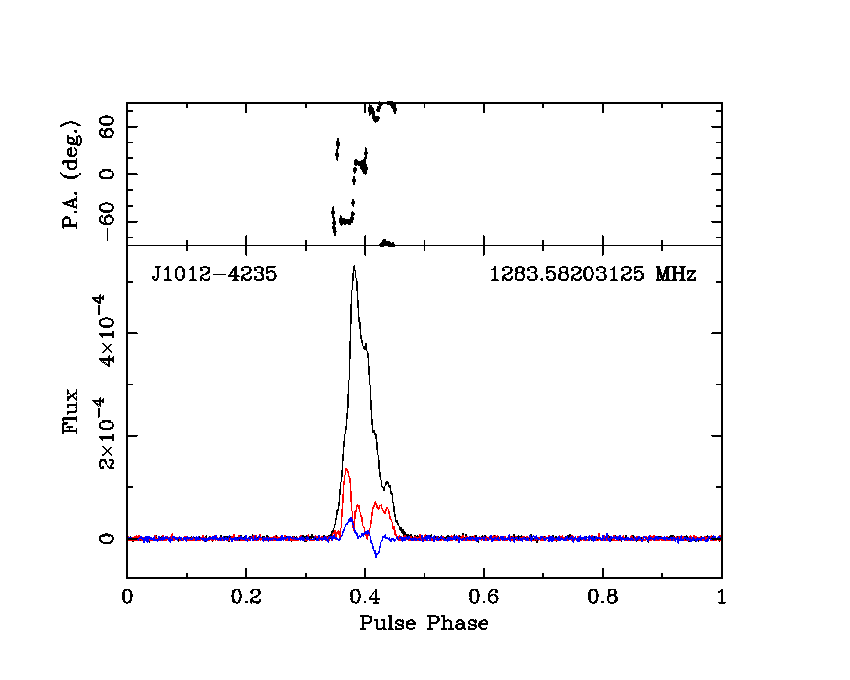}
        \caption{Polarisation calibrated profile of PSR~J1012$-$4235 formed from MeerKAT L-band data. {\em Bottom panel}: Intensity in arbitrary units as a function of rotational phase. The red colour shows linear polarisation and the blue colour shows circular polarisation across the rotational phase. {\em Top panel}: Variation of the position angle of the linear polarisation with spin phase.}
        \label{fig:mkt-pol-profile}
\end{figure}

As for almost all MSPs with nearly circular orbits, the spin axis of the pulsar is likely to be aligned with the orbital angular momentum; the reason for this is that the pulsar was spun up from orbiting material, and there has been no violent event to change the orbital plane \citep{2023pbse.book.....T}. The high inclination angle of the system (see discussion in Section~\ref{sec:timing_analysis})
therefore implies that our line of sight to the system is nearly perpendicular to the spin axis, which suggests the possibility that the pulsar is an orthogonal rotator. In such cases, we should expect to see emission from both magnetic poles in the form of an inter-pulse about $180\deg$ away from the main pulse.

 Figure~\ref{fig:mkt-pol-profile} shows the polarisation-calibrated profile of this pulsar created by integrating all the 73 observations covering a total of 13.2 hr of MeerKAT L-band data. 
A description of the polarisation calibration of the MeerKAT data of this pulsar is provided by \cite{2022PASA...39...27S}; the linear polarisation was derotated using the pulsar's well-known rotation measure (RM) of 61.80 rad m$^{-2}$.

Only 11\% of the total flux is polarised, of which the majority is linearly polarised, with a fractional linear polarisation of L/I = 23\%, and a fractional circular polarisation of V/I = 1.7\% (|V|/I = 6.3\%). The circular polarisation reverses its sign exactly at the centre of the on-pulse region. The position angle (PA) swing (see top panel of Fig.~\ref{fig:mkt-pol-profile}) shows steep transitions, which are likely caused by changes between orthogonal polarisation modes (OPMs).
As we can see in Fig. 2, all flux is concentrated in a narrow region, with no additional low-level components detectable anywhere else. After reducing the detection threshold for total intensity and polarised components, we can confirm that we do not detect any signature of the expected inter-pulse, which would be seen as emission at both poles. This refutes the idea of a simple orthogonal rotator.

\begin{figure}
\centering
        \includegraphics[width=\columnwidth]{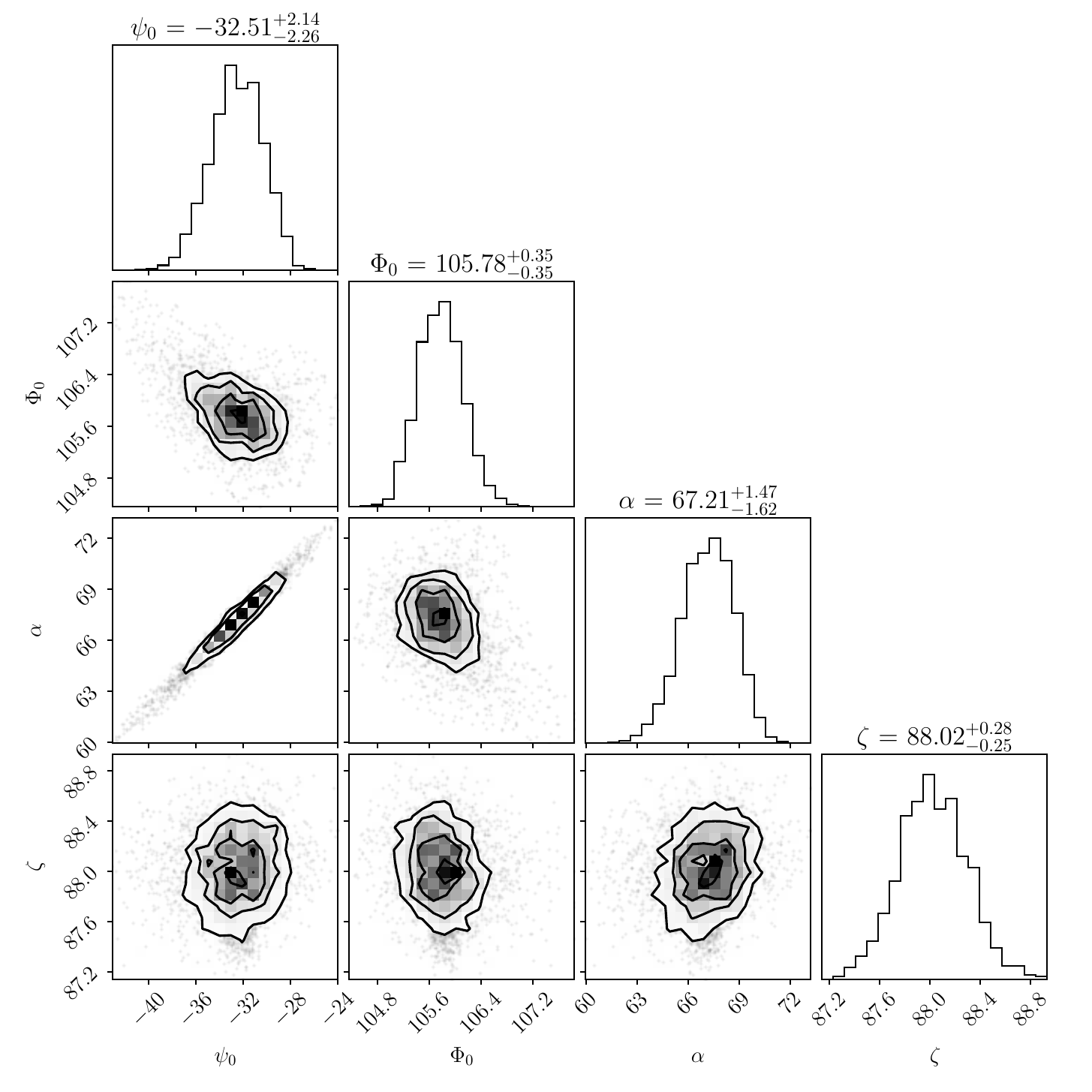}
        \caption{Corner plot with the parameters of the RVM fit to the polarisation profile of PSR~J1012$-$4235. $\Phi_0$ is the spin phase where the magnetic pole is closest to the line of sight, $\Psi_0$ is the position angle of the linear polarisation at $\Phi_0$, $\alpha$ is the angle between the spin axis and the magnetic axis, and $\zeta$ is the angle between the line of sight and the spin axis, which is given by $\zeta = \alpha + \beta$, where $\beta$ is the minimum angle between the line of sight and the magnetic axis.}
        \label{fig:rvm_corner}
\end{figure}

In order to gain some independent insight into the geometry of the system, we performed an RVM (see \citealt{1969ApL.....3..225R}) fit to the variation of the PA of linear polarisation with spin phase. Given the unknown complexities of the emission of radio pulsars, these results should be taken with some caution.

We performed the fit as described in \cite{2021A&A...654A..16G}, to which we refer
for further details. Briefly, we determined the geometry of the pulsar by fitting the RVM model using a Bayesian optimisation method as described by \cite{jk19} and \cite{2021MNRAS.504.2094K}. 
For this fit, we need four parameters, where the most important are the angle between the spin axis and the magnetic axis ($\alpha$), and the angle between the line of sight and the spin axis ($\zeta$), which is given by $\zeta = \alpha + \beta$, where $\beta$ is the minimum angle between the line of sight and the magnetic axis. The other two angles are $\Phi_0$, the spin phase where the magnetic pole is closest to the line of sight, and $\Psi_0$, which is the position angle of the linear polarisation at $\Phi_0$.
{\rm In our modelling, we constrain}  $\zeta$ by determination of the orbital inclination (from the assumption that the spin and orbital angular momenta are aligned). 
For the other parameters, we use uniform priors.
The code we used allows the possible existence of OPMs, that is,~ sudden 90 deg jumps in the PA values.
 The resulting parameters and their correlations are shown in Fig.~\ref{fig:rvm_corner}.

\begin{figure}
\centering
 \includegraphics[width=\columnwidth, trim={0 0 0 0},clip]{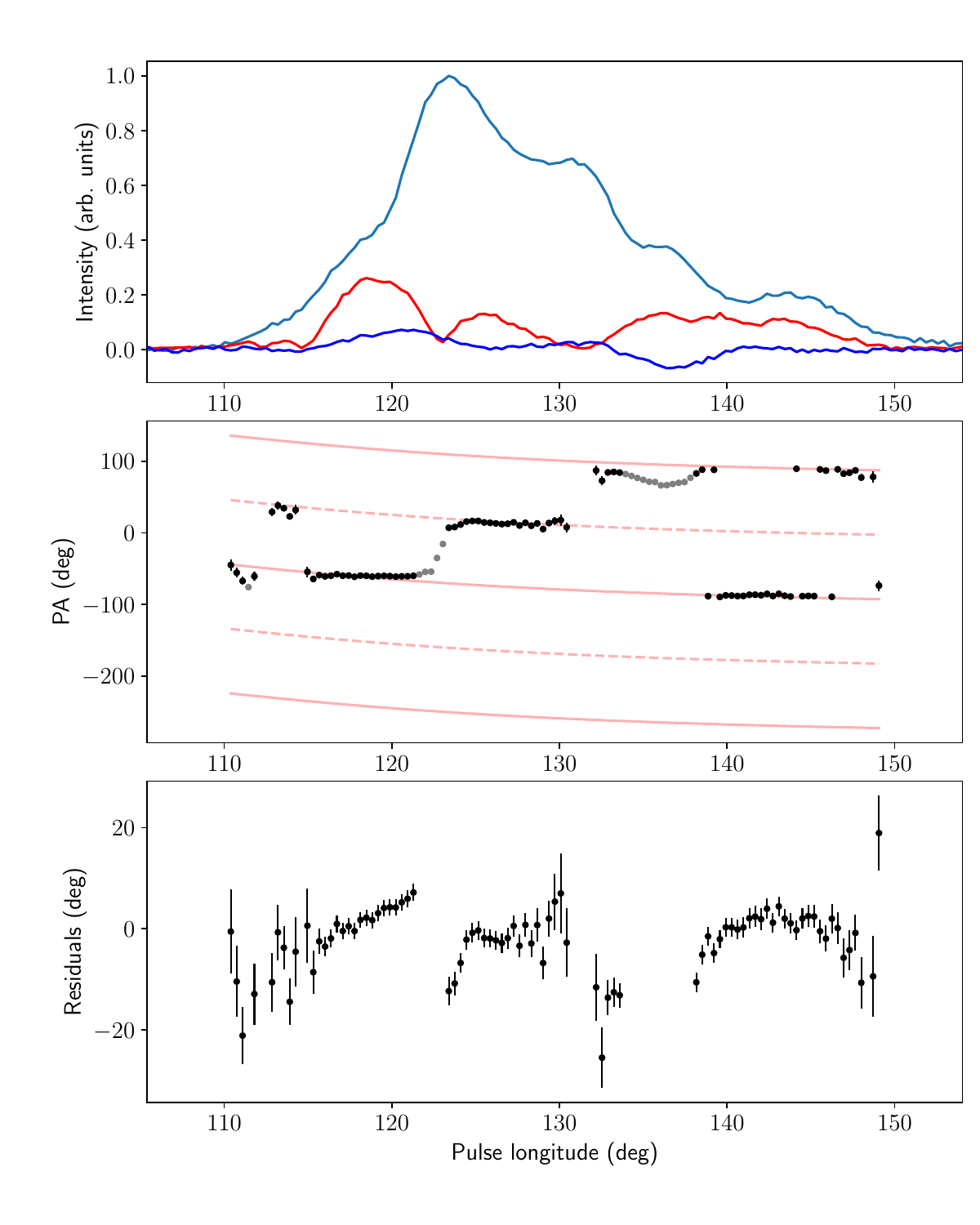}
        \caption{RVM fit with the polarised pulse profile.{\em Top plot}: Zoom into the region of the radio pulse of PSR~J1012$-$4235; we can see more clearly how the polarised emission changes with spin phase, or longitude. The colours are as in Fig.~\ref{fig:mkt-pol-profile}. {\em Middle plot}: Position angle as a function of the longitude. The black dots indicate the measured PA angles for the linearly polarised emission, and the smooth red curves (solid and dashed) indicate the prediction of the best-fit model. Polarisations that are $180 \deg$ apart are identical. Polarisations $90 \deg$ apart correspond to changes to OPMs.
    {\em Lower plot}: Differences between the measured PAs and the nearest model prediction accounting for possible OPM changes, i.e. the PA residuals. 
    }
        \label{fig:rvm_fit}
\end{figure}

In the middle panel of Fig.~\ref{fig:rvm_fit}, we can see how the prediction of this model compares with the observed PA of the linearly polarised emission as a function of spin phase, with the lower panel displaying the differences (residuals). As we can see, the observed PA follows the model reasonably well: the sudden $90 \deg$ changes in the PA are, as mentioned above, likely to be caused by changes of dominant OPMs. This notion is supported by the simultaneous drops in linear polarisation at those longitudes. Apart from these drops, the PA changes slowly with spin phase.

In the best-fit model, the reason for this slow change is that, even at closest approach, the magnetic axis passes far ($\beta \sim 20 \deg$) away from the line of sight. The fitted $\alpha \sim 67 \deg$ shows that this pulsar is not an orthogonal rotator (which would require $\alpha \sim 90 \deg$). This also means that we are looking at the edge of a wide ($> 20 \deg$) emission cone.  To some extent, this could explain why we see no inter-pulse: if the emission cone at the other magnetic pole is not as wide, we will not see radio emission from it.

However, even if the emission cone from the other magnetic pole is as wide as the cone from the pole we see, we might still miss it. This is because, in this system, the best-fit value of $\Phi_0$  at a longitude of $\sim 106 \deg$ is, unusually, {before} the start of the phase of strong radio emission. This geometry means that the earlier half of this emission cone has no associated emission; that is, it is clear that this emission cone is not fully illuminated. If this is the case, then the same might be happening in the parts of the emission cone from the other pole that cross our line of sight. 

In summary, although the orbital inclination is close to $90\, \deg$, the pulsar is not an orthogonal rotator, a condition that is common among the systems timed by RelBin \citep{2021MNRAS.504.2094K}. The fact that we see radio emission at $> 20 \, \deg$ from the magnetic field axis is common among MSPs, which generally have wide emission patterns (as shown by their generally long and complex pulse profiles). These broad emission patterns result in their large ($\sim 1$) beaming fractions.

The Shapiro delay measures only $\sin i$. This means that for each measurement of $\sin i$, there are two possible solutions for $i$ that are equidistant from edge-on ($90 \deg$).
We repeated the polarisation analysis for $\zeta \sim 92 \deg$, and obtain very similar results: in this case, we obtain $\alpha \sim 71 \deg$, which again results in $\beta \sim 20 \deg$. The quality of the fit is indistinguishable from the previous case ($\zeta \sim 88 \deg$), which means that the polarimetry data do not allow a choice between the two possible values of the orbital inclination.

\vspace{-1em}

\section{Timing analysis}

\label{sec:timing_analysis}

For the timing analysis, we used pulsar timing software \TEMPO\footnote{\url{http://tempo.sourceforge.net}}, because the specific orbital model we use in the analysis (i.e. ELL1H+, discussed in Section~\ref{subsec:binarymodels}) is only available in this timing package. The software first transfers all the ToAs from UTC to terrestrial time standard `TT(BIPM)', which is defined by the International Astronomical Union (IAU). We use NASA's JPL Solar System ephemeris, DE436 \citep{Folkner_Park+2016}, to account for the motion of the Earth.
For each ToA, the varying position of the radio telescope relative to the Solar System Barycentre (SSB) was then used to calculate the barycentric times of arrival. 
\TEMPO \, then calculates the phase residuals of each of the ToAs by comparing them to ToAs predicted by an initial timing model; we used the ephemeris presented in \cite{2015ApJ...810...85C} as our initial model.  
\TEMPO \, then adjusts the timing parameters to minimise the residual $\chi^2$. To compensate for the underestimated ToA uncertainties, which are possibly due to left-over systematic effects, we used separate weighting factors for each dataset; these are multiplied by their respective ToA uncertainties before the fit (see Table~\ref{table:observations_1012_details}). 

\subsection{Binary models}
\label{subsec:binarymodels}
To fit for the binary parameters, we used the ELL1H+ binary model \citep{2021A&A...654A..16G}, which is implemented in the latest versions of \texttt{TEMPO} (> 13.102). This is a theory-independent model based on the ELL1 model \citep{2001MNRAS.326..274L}, which is designed to avoid
the correlation between the epoch of periastron, $T_0$, and the longitude of periastron, $\omega, $ in low-eccentricity binaries by reparameterising the orbit and measuring the epoch of the ascending node, $T_{\rm asc}$, instead of $T_0$. 
Instead of fitting for $\omega$ and the orbital eccentricity $e$, this model fits for the Laplace-Lagrange parameters: $\epsilon_1 = e \cos \omega$ and $\epsilon_2= e \sin \omega$. 
In addition to the ELL1 model, the ELL1H+ model includes an extra term of the order of $xe^2$ in the expansion of the R\"omer delay \citep{2019MNRAS.482.3249Z}, which makes it applicable to most binary pulsars. As the next-order term for PSR~J1012$-$4235 ---namely ${\cal O} (xe^3) \sim 0.9\, $ns--- is much smaller than the uncertainty in our measurement of the $h_3$ parameter ($\sim 55$ ns), ignoring it does not affect our measurement of the Shapiro delay.

The ELL1 model measures the Shapiro delay in the form of two PK parameters, namely range ($r$) and shape ($s$), which according to GR are given by $r= T_\odot \Mc$ and $s=\sin i$ and where $T_{\odot} \equiv {\cal G M}^{\rm N}_{\odot}/c^3 = 4.9254909476412675...\, \upmu \rm s$ is an exact quantity, and represents the solar mass parameter (${\cal G M}^{\rm N}_{\odot}$, \citealt{Prsa2016}) in time units.
The ELL1H+ model instead uses the orthometric parameterisation of Shapiro delay \citep{2010MNRAS.409..199F}, which removes the high correlation between the $r$ and $s$ parameters, leading to a robust quantification of this effect. The Shapiro delay parameters are the orthometric amplitude, $h_3$, and the orthometric ratio, $\varsigma,$ which in GR are given by:
\begin{equation}
    \varsigma = \frac{\sin i}{1 + \sqrt{1 - \sin^2i}}, \,
    h_3= T_\odot \Mc \varsigma^3.
\end{equation}
To fit for these parameters, the ELL1H+ model uses the exact expression for the Shapiro delay, that is, equation (29) in \cite{2010MNRAS.409..199F}.


\subsection{Results from radio and gamma-ray timing}
\label{sec:results_timing}
\begin{figure*}
  \centering
  \subfloat[][]{\includegraphics[scale=0.7]{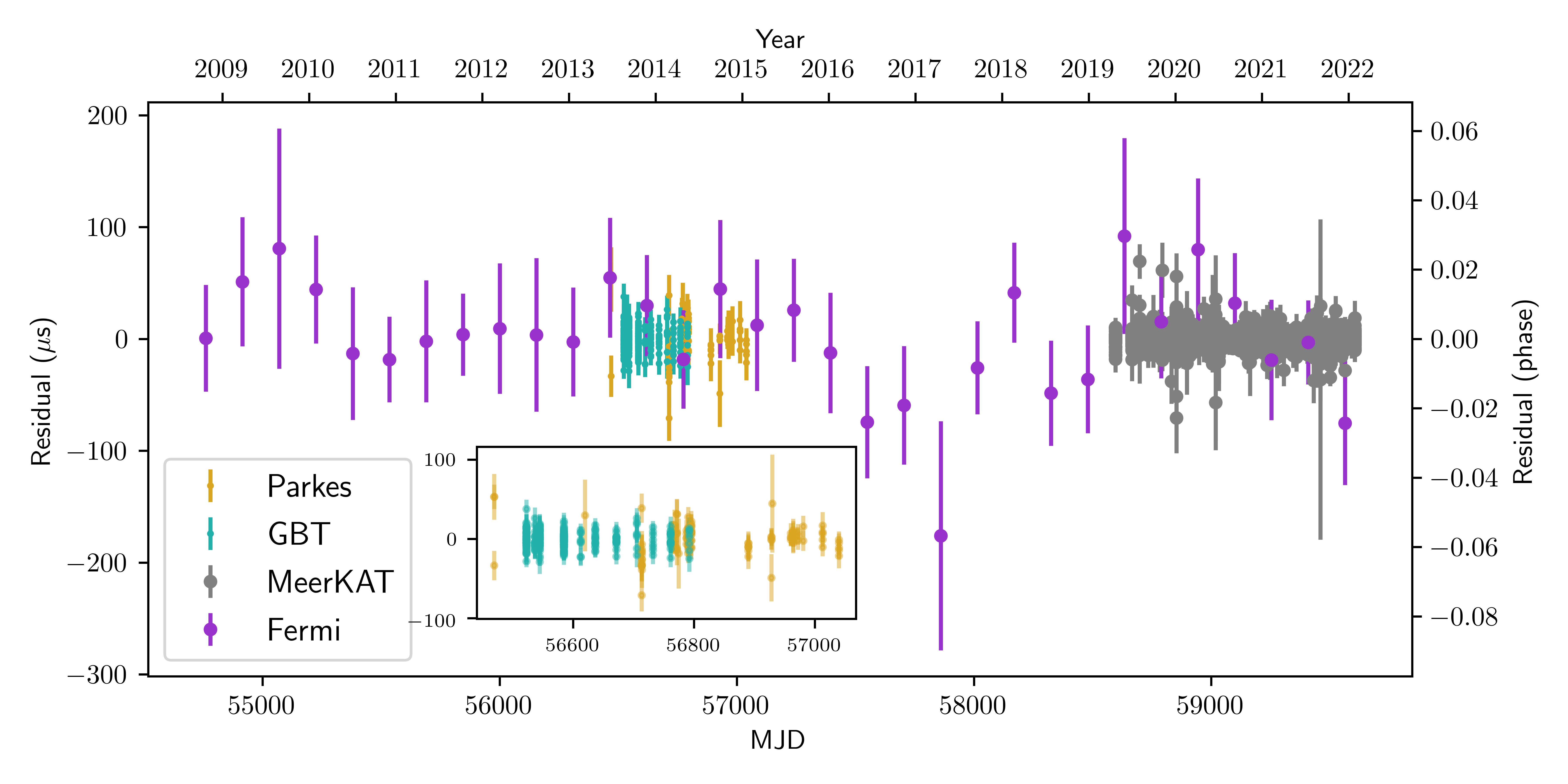} \label{fig:residual-plot-1012} } \\
  \subfloat[][]{\includegraphics[scale=0.7]{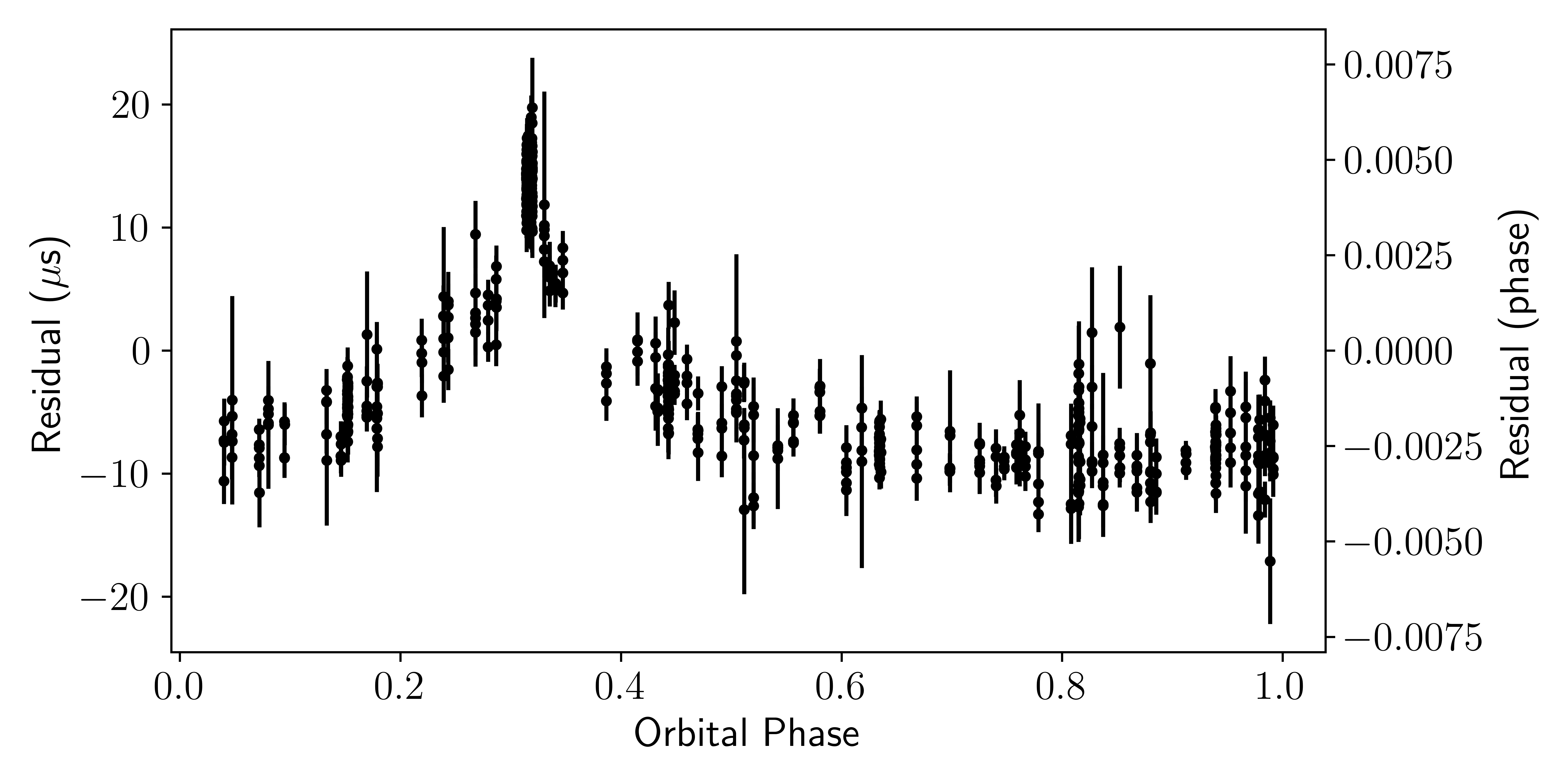} \label{fig:residual-plot-shapiro}}
  \caption{Residuals obtained from the fit of the timing model for PSR~J1012$-$4235. \textit{Top plot:} Post-fit residuals  showing ToAs from Parkes, GBT, MeerKAT, and Fermi-LAT data, with error bars representing 1$\sigma$ uncertainty in ToAs. \textit{Bottom plot:} Residuals of MeerKAT ToAs, excluding the fit for Shapiro delay effect but fixing all the other system and orbital parameters obtained from the full fit. The orbital phase is measured from periastron and the peak in the Shapiro delay signature occurs around the phase of superior conjunction of the pulsar. \label{fig:residual-plots-1012}}
\end{figure*}

From the fit of the ELL1H+ binary model to all the radio and $\gamma$-ray data, we derived a phase-connected timing solution for this system (see Table~\ref{tab:timing_solutions_1012}). This is a unique solution for the binary that correctly predicts the pulse arrival times for every rotation of the pulsar in the data set. Figure~\ref{fig:residual-plot-1012} shows the post-fit ToA residuals as a function of time. The addition of Fermi-LAT ToAs (detailed in Section~\ref{sec:gamma-data}) is especially useful in precisely constraining the long-term astrometric parameters for the system. We get a reduced $\chi^2$ of 1.0084 after adding the EFAC parameter. The resulting residuals show a Gaussian distribution with no trends (or structure), indicating that there are no detectable unmodelled effects.

The resulting best-fit solution includes precise astrometric, spin, and orbital parameters. The astrometric parameters include a precise proper motion in right ascension (RA) and declination (DEC) directions, resulting in a total heliocentric proper motion ($\mu_{\rm T}$) of 6.58(5) mas yr$^{-1}$, as well as a parallax of 0.85(50) mas.
This measurement is important, because it excludes one of the distance estimates of the YMW16 model (see~\ref{sec:discussion_1012}).
The spin parameters include two significant DM derivatives and one spin frequency derivative.

Regarding the orbital parameters, we find a strong detection of Shapiro delay with $h_3 = 1.222(54) \upmu$s and $\varsigma = 0.9646(49)$. The variation of the Shapiro delay with orbital phase can be clearly seen in Fig.~\ref{fig:residual-plot-shapiro}. This detection implies a companion mass of $\Mc=0.276$(16) \msun, and a $\sin i = 0.99937(18)$, meaning an inclination $i$ of $87.97^{+0.31}_{-0.27}\deg$. The pulsar mass ($\Mp$) derived from these measurements using the mass function is 1.44(13) \msun. Our measurement of the companion mass aligns well with the $\sim$ 0.27 \msun \, predicted by the $P_{\rm orb}$-- M$_{\rm WD}$ relationship of \cite{1999A&A...350..928T} for He WDs.
The eccentricity of 3.45$\times 10^{-4}$ agrees well with the theoretical range predicted by \cite{1994ARA&A..32..591P} for a MSP--WD binary with an orbital period of 37.9 days.

Regarding the measurement of the advance of periastron, $\dot{\omega,}$  in this binary, its prediction from GR gives a value of $6.5 \times 10^{-4}$ $\deg \rm yr^{-1}$, while our fit for this parameter yields $(1.4 \pm 1.5) \times 10^{-3}$ $\deg \rm yr^{-1}$. Although the measurement is consistent with the expected value, its uncertainty is still
about two times larger than the expected effect, which means that we cannot yet measure this effect. However, measuring this parameter should be feasible in the near future with continued timing with MeerKAT observations.

\begin{table}

\caption{Phase-connected timing solution for PSR J1012$-$4235, making use of combined radio and $\gamma-$ray data.}
\label{tab:timing_solutions_1012}
\begin{center}{\footnotesize
\setlength{\tabcolsep}{0.5pt}
\renewcommand{\arraystretch}{1.2}
\begin{tabular}{l c}
\hline
\multicolumn{2}{c}{Parameters of timing solution} \\
\hline\hline
Terrestrial Time Standard                                             \dotfill &  TT(BIPM2019)                                                        \\
Time Units                                                            \dotfill \dotfill   & TDB                                                             \\
Solar System Ephemeris                                                \dotfill \dotfill &  DE436                                                           \\
Reference Epoch (MJD)                                                 \dotfill 
& 59111.238529  
                          \\
Start of Timing Data (MJD)                                            \dotfill    & 54760.138                                                         \\
End of Timing Data (MJD)                                              \dotfill & 59604.959                                                        \\
Number of ToAs                                                        \dotfill  & 3971                                                     \\
Residual RMS ($\upmu$s)                                                \dotfill        & 3.527                                           \\
 $\chi^2$                                                               \dotfill        & 3983.29
 \\
 Reduced $\chi^2$                                                       \dotfill        & 1.0084                                   \\
\hline
\multicolumn{2}{c}{Astrometric and spin parameters}  \\
\hline\hline
Right Ascension, $\alpha$ (J2000)                                     \dotfill &             10:12:12.9388541(46)                          \\
Declination, $\delta$ (J2000)                                         \dotfill &        $-$42:35:53.403185(57)                                   \\
Proper Motion in $\alpha$, $\mu_\alpha$ (mas yr$^{-1}$)               \dotfill    & $-$4.093(49)                                                       \\
Proper Motion in $\delta$, $\mu_\delta$ (mas yr$^{-1}$)               \dotfill 
& 5.154(52)        
                                   \\
Parallax, $\varpi$ (mas)      \dotfill 
 
& 0.85(50) 
\\
Spin Frequency, $f$ (Hz)                                        \dotfill    & 322.4619326648837(30)                                           \\
1st Spin Frequency derivative, $\dot{f}$ (Hz s$^{-1}$)                \dotfill        & $-$6.8062(8)$\times 10^{-16}$                             \\

Dispersion Measure, DM (pc cm$^{-3}$)                                 \dotfill     &     
71.650839(77)                                                     \\
1st derivative of DM, DM1 (pc cm$^{-3}$ yr $^{-1}$)
     \dotfill & 
     
     0.000922(88)  \\
     2nd derivative of DM, DM2 (pc cm$^{-3}$ yr $^{-2}$)
     \dotfill &
     0.000547(36)  
 \\
Rotation Measure (rad m$^{-2}$)     \dotfill & 61.80 \\
\hline
\multicolumn{2}{c}{Binary Parameters}  \\
\hline\hline
Binary Model                                                          \dotfill     & ELL1H+                                                       \\
Epoch of ascending node, $T_{\rm asc}$ (MJD)                           \dotfill                 & 58862.548525754(33)                                  \\
Orbital Period, $P_b$ (days)                                          \dotfill        &          37.9724631823(24)                               \\
Projected Semi-major Axis, $x_{\rm p}$ (lt-s)                         \dotfill  &          21.26306899(30)                                    \\
$\epsilon_1$                                                          \dotfill &     $-$0.000140974(16)\\
$\epsilon_2$                                                          \dotfill &     0.0003156766(88)  \\

    
 $h_3$  ($\upmu$s)      \dotfill & 1.222(54) \\
 $\varsigma$     \dotfill & 0.9646(49) \\

\hline
\end{tabular} }
\end{center} 

\end{table}

\begin{table}
\caption{Derived astrometric, spin, and orbital parameters for PSR~J1012$-$4235.}
\label{tab:derived_parameters}
\begin{center}{\footnotesize
\setlength{\tabcolsep}{4pt}
\renewcommand{\arraystretch}{1.2}
\begin{tabular}{l c c}
\hline
\multicolumn{2}{c}{Derived Parameters}  \\
\hline\hline
Galactic longitude, $l$ (deg) \dotfill & 274.2176 \\
Galactic latitude, $b$ (deg)  \dotfill & 11.2246  \\
DM derived distance$-$NE2001 model (kpc)       \dotfill  & 2.51  \\
DM derived distance$-$YMW16 model (kpc)       \dotfill  & 0.37  \\
Distance from timing parallax (kpc)     \dotfill & 1.18$^{+1.68}_{-0.44}$\\
Composite proper motion $\mu_{\rm T}$ (mas yr$^{-1}$)       \dotfill & 6.582(51)   \\ 
Heliocentric transverse velocity, $v_{\rm T}$ (km s$^{-1}$) \dotfill  & 36.8$^{+52.4}_{-13.7}$ \\
\hline
Spin Period, $P$ (s)                                              \dotfill &  0.00310114124707943(3)                             \\
$\dot{P}_{\rm obs}$  (10$^{-21}$ s s$^{-1}$)     \dotfill  &  6.5456(7) \\
$\dot{P}_{\rm kin}$  (10$^{-24}$ s s$^{-1}$)$^a$     \dotfill  &  $6.0^{+91.4}_{-7.4}$ 
\\
Intrinsic Spin-down, $\dot{P}_{\rm int}$ (10$^{-21}$ s s$^{-1}$)      \dotfill  &  $6.54^{0.007}_{-0.09}$                                                                \\
Surface Magnetic Field, $B_s$ (10$^{8}$ G)                           \dotfill & 1.44 
\\
Characteristic Age, $\tau_{\rm c}$ (Gyr)                              \dotfill  & 7.50 
\\
Spin-down power, $\dot{\rm E}$ ($10^{33} \rm erg \, \rm s^{-1}$) \dotfill & 8.68 
\\
\hline
Epoch of Periastron, $T_0$ (MJD)                           \dotfill & 58897.98269(27)                              \\
Orbital Eccentricity, $e$                                             \dotfill   &   3.4572(1)$\times 10^{-4}$                          \\
Longitude of Periastron, $\omega$ (deg)                               \dotfill      &                335.9355(25)                              \\
Mass Function, $f(M_{\rm p})$ (${\rm M}_\odot$)                       \dotfill   &               0.0071585169(3)                                \\     
Companion Mass, $M_{\rm c}$ (M$_\odot$)                               \dotfill    & 0.276(16)       \\
$\sin i$   \dotfill                 &   0.99937$^{+0.00015}_{-0.00018}$  \\
Pulsar Mass, $\Mp$ (M$_\odot$)        \dotfill & 1.43(13) \\
Total Mass, $\Mtot$ (M$_\odot$)                                           \dotfill &   1.710(149)                               \\
 $\dot{P}_{\rm b, \rm kin}$ (10$^{-14}$ s s$^{-1}$) $^b$ \dotfill & $0.64^{+9.67}_{-0.78}$ 
\\
$\dot{P}_{\rm b,GR}$ ($10^{-17}$ s s$^{-1}$) $^c$ \dotfill & $-1.68^{+0.20}_{-0.21}$\\ 
\hline
\end{tabular} }
\end{center} 
\footnotesize{$^a$Total kinematic contribution to spin period derivative.\\$^b$Total kinematic contribution to orbital period derivative.\\$^c$GR contribution to orbital period derivative.}
\end{table}

\subsection{Mass constraint from the Shapiro delay detection}

\begin{figure}
\centering
\subfloat{
    \includegraphics[width=\linewidth]{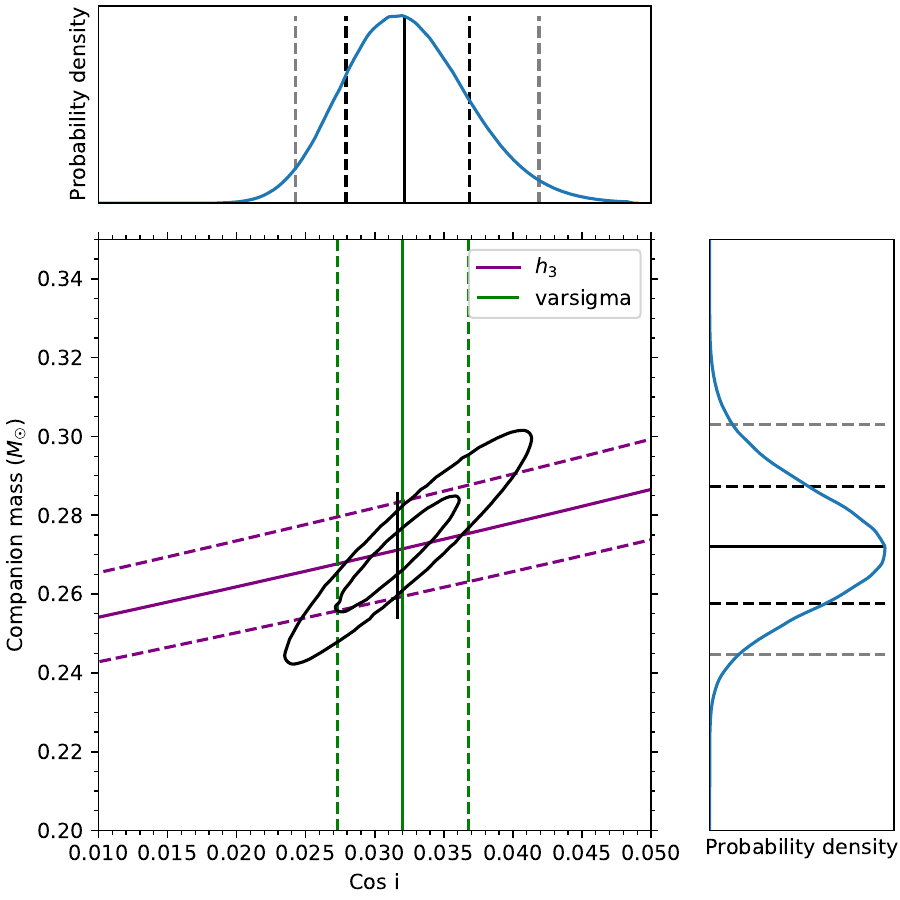}}
\hfill
\subfloat{
    \includegraphics[width=\linewidth]{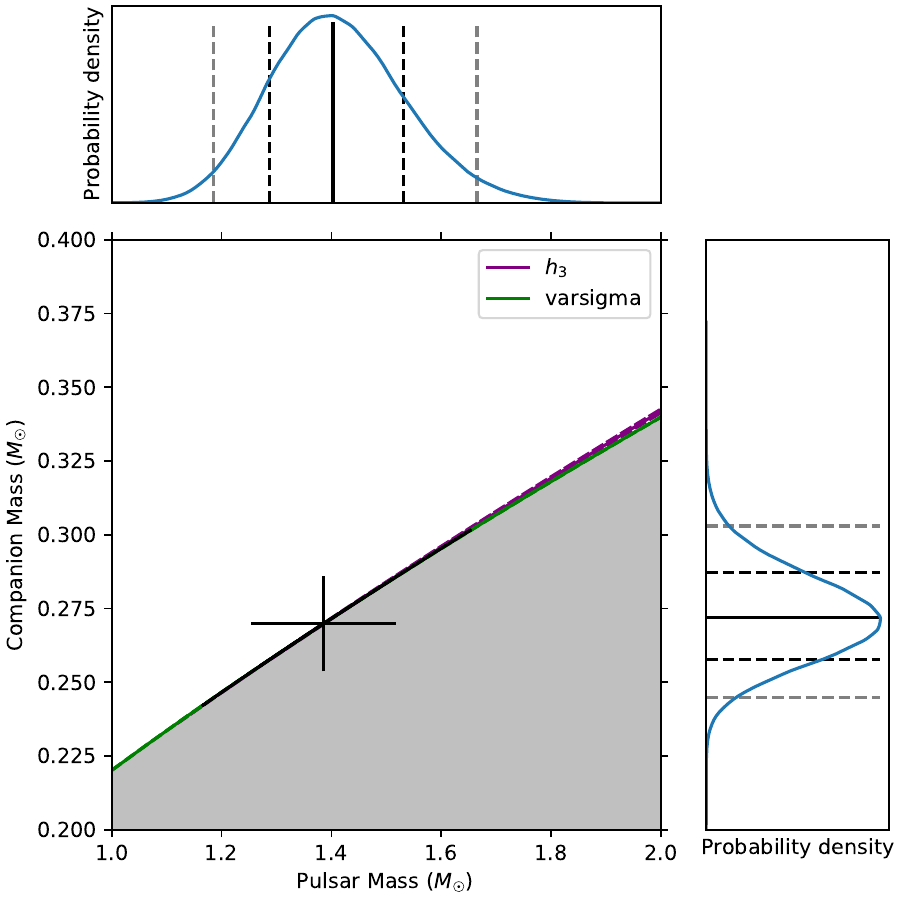}}

\caption{Constraints on $\Mc$, $M_{\rm p}$, and $\cos i$ derived from the measurement of Shapiro delay. Coloured dashed lines depict the 1$\sigma$ uncertainty on the measurement. Black contours show the likelihood or 2D PDFs at 39 $\%$ (inner) and 85 $\%$ (outer) confidence ellipse from the $\chisquare$ maps on ($\cos i-\Mc$) and ($\Mp-\Mc$) grids derived from the ELL1 binary model fit. The corner plots represent the marginalised 1D PDFs of each of these measurements and the vertical lines on these mark the median, 1$\sigma$, and 2$\sigma$ estimates. The black crosses indicate the best masses fit by the DDGR solution (a theory-dependent variation of the DD binary model, \citealt{1987grg..conf.....M,1989ApJ...345..434T} which assumes GR) and the best inclination derived from the mass function using those masses.}
\label{fig:mp-mc-cosi-1012-planes}
\end{figure}

In order to better understand the correlations between the Shapiro delay parameters and better estimate their uncertainties and the uncertainties on the masses and orbital inclination, we performed a Bayesian study of the masses and orbital inclination of the system. First, we created a uniform grid of points on the ($\cos i - \Mc$) plane and derived the best-fit $\chi^2$ values from the TEMPO fit of this model on each of the points in this grid. Second, these $\chi^2$ values are then converted to a Bayesian likelihood using the relation by \cite{2002ApJ...581..509S}: $p(\cos i, M_{\rm c}) \propto e^{\frac{\chi^2_{\rm min} - \chi^2}{2}}$, where $\chi^2_{\rm min}$ is the minimum $\chi^2$ value. These likelihoods are then used in the calculation of joint posterior probability density functions (PDFs), both in the original plane and also in the ($\Mp - \Mc$) plane.

Figure~\ref{fig:mp-mc-cosi-1012-planes} shows the Shapiro delay parameters derived from the ELL1H+ model on the ($\cos i - \Mc$) and ($\Mp - \Mc$) planes. Black contours represent the 39$\%$ (inner) and 85$\%$ (outer) confidence ellipses on the joint posterior PDFs, which represent the 1$\sigma$ and 2$\sigma$ error bars in 1D PDFs. Using this method, we get $\cos i = 0.0339^{+0.0049}_{-0.0044}$, implying an inclination of 88.06$^{+0.28}_{-0.25} \deg$, $\Mc = 0.270^{+0.016}_{-0.015}$ \msun, and $\Mp = 1.44^{+0.13}_{-0.12}$ \msun, with errors representing 1$\sigma$ uncertainties. These values and uncertainties agree very well with those derived in Section~\ref{sec:results_timing}. 

\subsection{Kinematic effects on the spin and orbital period}
\label{sec:discussion_1012}
\begin{figure}
\centering
        \includegraphics[width=1.1\linewidth]{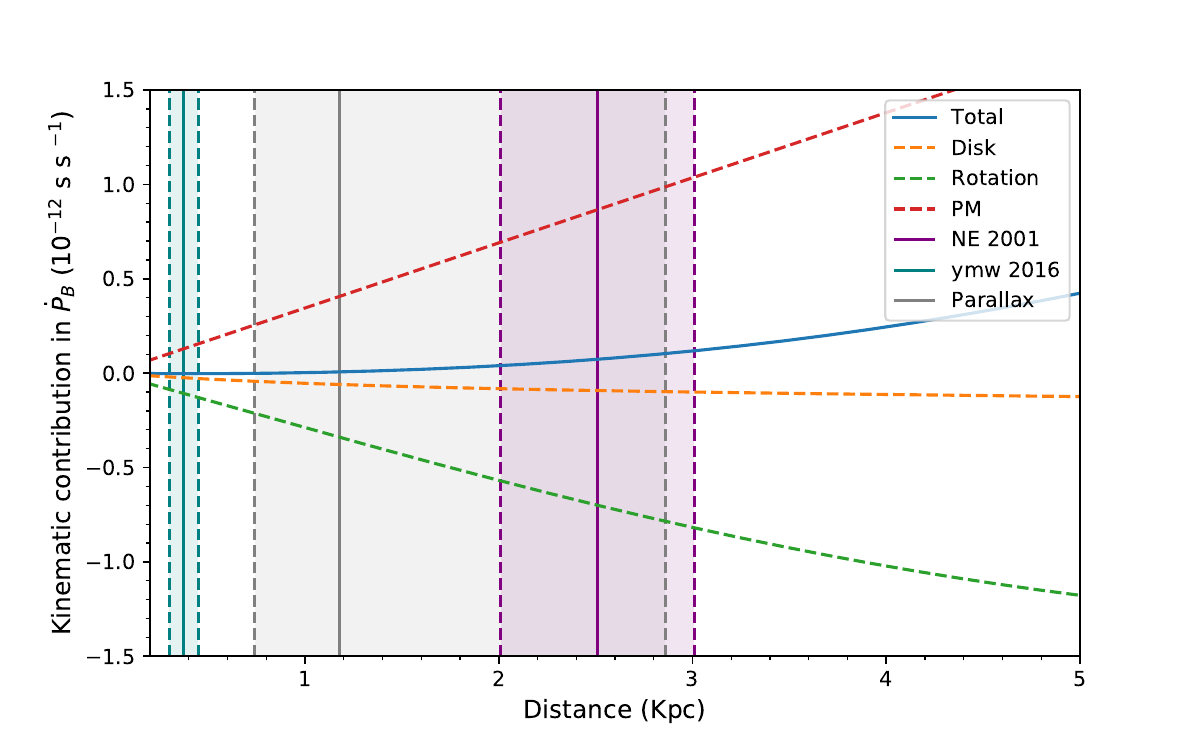}
        \caption{Acceleration contributions to the orbital period derivative due to kinematic effects as a function of distance. Orange and green curves show the vertical and differential accelerations from the Galactic disc, while the red curve shows the contribution from the proper motion of the pulsar. The blue curve represents the total acceleration contribution. A distance uncertainty of 20$\%$ is assumed on the DM distance predictions from the NE2001 model (purple shaded region) and the YMW16 model (teal shaded region), while the 1$\sigma$ uncertainty is shown for the parallax measurement from timing (grey shaded region).} 
        \label{fig:pbdot-dist-1012}
\end{figure}

The two most broadly accepted Galactic electron density models are the models YMW16 \citep{2017ApJ...835...29Y} and NE2001 \citep{2002astro.ph..7156C}. These models can predict the distance to the pulsar based on the line-of-sight electron density and the DM of the pulsar. For this pulsar, YMW16 predicts a distance of 0.37 kpc, while NE2001 predicts 2.51 kpc. These estimates are generally thought to be accurate to $\sim 20\%$, but in this case it is clear that they do not agree at this level (see Fig.~\ref{fig:pbdot-dist-1012}). However, in the timing solution, we fit for a parallax and obtain 0.85(50) mas, indicating a distance ($d = 1 /\varpi$) of 1.18$^{+1.68}_{-0.44}$ kpc, a value in between the predictions from both the models but more aligned with NE2001 model within 1$\sigma$. 

The observed spin period derivative of the pulsar ($\dot{P}_{\rm obs}$) includes contributions not only from the intrinsic spin period derivative but also from the derivative of the Doppler factor. The later includes two terms, representing (a) the proper motion of the pulsar,
an effect known as the Shklovskii effect (the first term in the equation below) and (b) the Galactic acceleration \citep{2012hpa..book.....L}:
\begin{equation}
    \frac{\Pdotobs}{P} = \frac{\Pdotint}{P} + \frac{\mu_{\rm T}^2d}{c} + \frac{a_{\rm gal,disc}}{c} + \frac{a_{\rm gal, rot}}{c},
\end{equation}
where $\Pdotint$ is the intrinsic spin period derivative, $a_{\rm gal,disc}$ is the vertical acceleration contribution from the Galactic disc, and  $a_{\rm gal,rot}$ is the acceleration due to the differential rotation of the galaxy.

These Galactic accelerations are calculated from the analytical expressions given by \cite{1991ApJ...366..501D, 1995ApJ...441..429N,2009MNRAS.400..805L}:
\begin{eqnarray}
     a_{\rm gal, rot} & = & -\frac{\Theta_0^2}{R_0 c}\left(\rm cos \emph{l}+\frac{\beta}{\beta^2+\rm sin^2\emph{l}}\right)\rm cos \, \it{b} ,\\
     \frac{a_{\rm gal, disc}}{10^{-9}\, \rm cm\, s^{-2}}& = & \left(2.27 z_{\rm kpc} + 3.68 (1-e^{-4.3z_{\rm kpc}})\right) |\rm sin\, \it{b}| ,
\end{eqnarray}
where $\beta \equiv \left({d/R_0}\right)\rm cos \emph{b}- \rm cos\emph{l}$ and $z_{\rm kpc} \equiv |d \rm sin\emph{b}|$ in kpc. $R_0$ is the distance to the Galactic centre, 8.275(34) kpc, and $\Theta_0$ is the Galactic rotation velocity, 240.5(41) km s$^{-1}$ (values taken from
\citealt{2021A&A...654A..16G}). This approximation is valid given the fact that the pulsar is relatively nearby and likely has a low Galactic height. 

Assuming the distance predicted by our parallax measurement, 
we obtain an acceleration of $0.58^{+8.9}_{-0.7} \times 10^{-12} \rm m s^{-2}$  for the sum of kinematic corrections. 
Using this, we get the value of kinematic correction to $\dot{P}$, $\dot{P}_{\rm kin} = 6 \times 10^{-24}$. Thus
$\Pdotint = 6.54^{0.007}_{-0.09} \times 10^{-21} \rm s\,s^{-1}$, 
which is very close to the $\Pdotobs$ because $\dot{P}_{\rm kin}$ is 100-1000 times smaller. With this estimate of $\Pdotint$, we estimate the surface magnetic field, $\rm B_s$, the characteristic age, $\tau_c$, and the spin-down luminosity of the system, $\dot{E,}$ of the pulsar using the equations in \cite{2012hpa..book.....L}. Table~\ref{tab:derived_parameters} presents the estimates for the derived parameters along with their 1$\sigma$ uncertainties.

 Similar kinematic effects to those discussed above affect the measurement of the observed orbital period derivative, $\Pbdotobs$, as well.
 Assuming the distance from the parallax, we estimate the total kinematic contribution $\dot{P}_{\rm b, kin} =  0.64^{+9.67}_{-0.78}\times 10^{-14} \rm s \, s^{-1}$. 
The expected orbital decay due to the quadrupolar GW emission from this system is $-1.68^{+0.20}_{-0.21}\times 10^{-17} \rm s \rm s^{-1}$ (assuming GR). The observed $\dot{P}_{\rm b}$ is $1.4 \pm 4.3\times10^{-12} \rm s \rm s^{-1}$. Although this measurement is consistent with expectations, its uncertainty is still two orders of magnitude higher than the expected contributions from kinematic effects and is nearly five orders of magnitude larger than the orbital decay from GW emission. Therefore, we cannot yet constrain the kinematic effects from timing. 
  
Figure~\ref{fig:pbdot-dist-1012} shows each of the contributions to $\Pbdotobs$ from the kinematic effects as a function of distance.  
As we can see, the total contribution to $\dot{P}_{\rm b}$ from the kinematic effects (blue curve) is extremely flat. This turns out to be one of the most interesting features of this system. Even with the relatively large distance uncertainty, this contribution is not only small but it has a small associated uncertainty, $\delta \dot{P}_{\rm b,kin}= 9\times 10^{-14}$ s s$^{-1}$. This has an important consequence: it means we will be able to obtain a tight constraint on the variation of the gravitational constant $\dot{G}/G$ \citep{1993tegp.book.....W, 2011LRR....14....2U}.
Furthermore, this will happen relatively soon, because the precision of $\dot{P}_{\rm b, obs}$ will continue to improve, and relatively quickly, with the relative error scaling as $T^{-5/2}$, where $T$ is the timing baseline; it will, in reality, improve even faster because the precise MeerKAT observations started only very recently. This will result in improved estimates of the intrinsic $\dot{P}_{\rm b}$, which will only be limited by the  $\delta \dot{P}_{\rm b,kin}$. The constraint on $\dot{G}/G$ will then be proportional to $\delta \dot{P}_{\rm b,kin}/P_{\rm b} = 8.6 \times 10^{-13}\, \rm yr^{-1}$. 

\section{Gamma-ray and radio pulse profiles}
\label{sec:gammaandradio}


With our new timing solution obtained from radio data, we refolded the $\gamma-$ray data to create a $\gamma-$ray pulse profile.  We use a single phase-connected solution on both the radio and $\gamma-$ray data such that the profiles are phase-aligned; that is, the start of the rotational phase indicates the same reference epoch.

\begin{figure}
\centering
        \includegraphics[width=0.9\linewidth]{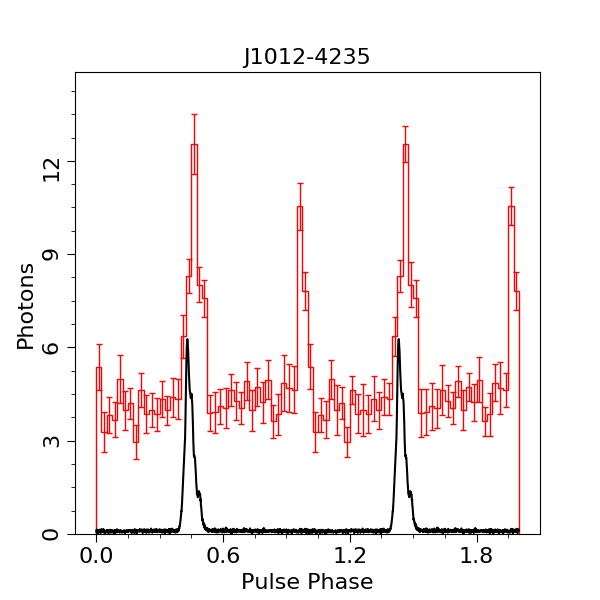}
        \caption{Comparison of the $\gamma-$ray pulse profile (red) with the radio profile (black) of PSR J1012$-$4235. The profile is formed using a phase-connected solution (ephemeris) such that both the folds take the same reference epoch as the start of the profile.}
        \label{fig:radio-gamma-profile}
\end{figure}

Figure~\ref{fig:radio-gamma-profile} shows a comparison of the two profiles. Unlike in radio, we do see a second peak in the pulsed emission, which is superficially consistent with the previously mentioned idea of the orthogonal rotator. However, like the radio polarisation, this gamma-ray profile does not support the idea of an orthogonal rotator: the pulse morphology seen here is similar to that of most $\gamma$-ray pulsars \citep{2023arXiv230711132S}, where a
single radio pulse, after a small offset in phase (which is visible
here), is followed by two strong peaks in the $\gamma$-ray emission, linked by a bridge of emission, which is also faintly visible here. 

This is clear evidence supporting the idea that the radio and $\gamma$-ray emission do not generally originate from the same region of the pulsar magnetosphere. There are a few exceptions, such as the first MSP (PSR~B1937+21) and the first `black widow' pulsar, PSR~B1957+20, where both the radio and $\gamma$-ray emission show a precisely co-located pulse and inter-pulse (for a detailed discussion on these topics, see e.g., \citealt{2012ApJ...744...33G}). The radio emission of those systems is more consistent with that of an orthogonal rotator.
Further analysis in an attempt to constrain the pulsar magnetosphere using $\gamma-$ray data will be carried out in the future.


\section{Conclusion}
\label{sec:conclusion_1012}

In this paper, we present the results of our timing analysis of PSR J1012$-$4235, which we performed with data obtained with the Parkes (1.5 years), GBT (7 months), and MeerKAT (2.7 years) radio telescopes, and the Fermi $\gamma$-ray space telescope (13 years), covering a total time baseline of 13 years. We present the phase-connected timing solution of this pulsar, including refined estimates of astrometric, kinematic, and orbital parameters. We measured the proper motion of the pulsar to be 6.5 mas yr$^{-1}$. With the help of dense MeerKAT observations that cover the superior conjunction of the pulsar, we detect a significant Shapiro delay signature in the binary, the first relativistic effect in this system. We obtain a 22$\sigma$ detection of the $h_{3}$ parameter of 1.222(54) $\upmu$s and a $\sim$ 200$\sigma$ detection of $\varsigma = 0.9646(49)$. This yields measurements of the component masses and the orbital inclination: $\Mp=1.44^{+0.13}_{-0.12}$ \msun, $\Mc = 0.270^{+0.016}_{-0.015}$ \msun , and $i=88.06^{+0.28}_{-0.25} \deg$.

Using the phase-connected solution, we also phase-aligned the radio- and $\gamma-$ray profiles of the pulse. This alignment shows that the pulse peak of the radio profile precedes the  peak in the $\gamma$ profile, which in turn aligns with one of the features of the radio profile.

Lastly, with the proper motion and parallax measurements from timing, we constrained the kinematic contributions to the observed spin period and orbital period derivatives. We note that the curve for the contribution of the kinematic effects to $\Pbdotobs$ is nearly flat within the range of distances allowed by our measurement of the parallax. Therefore, despite the large uncertainty in the distance measurement, we obtain a very small uncertainty in the kinematic contributions to $\Pbdot$. This represents the limiting factor for estimating the intrinsic $\Pbdot$ of the system from future timing measurements. Therefore, continued timing of this system might eventually result in a limit on $\dot{G}/G$ of $8.6 \times 10^{-13}\, \rm yr^{-1}$. For PSR J1713+0747, which currently provides the best limit on this parameter, this parameter is 8.1$\times10^{-13}\, \rm yr^{-1}$ \citep{2019MNRAS.482.3249Z}. Thus, the improvement of the $\dot{P}_{\rm b}$ measurement of PSR~J1012$-$4235 has the potential to provide a limit on $\dot{G}/G$ comparable to that provided by PSR~J1713+0747.

However, with continued timing, the measurement of timing parallax will improve and therefore the distance uncertainty will decrease in future. 
If the distance to the pulsar is near the best current estimate, then a two-fold improvement in the measurement of the parallax (with $\delta \rm \varpi \sim 0.25 mas$) will result in an order-of-magnitude improvement in the uncertainty of $\dot{P}_{\rm b,kin}$ for this system, which will mean a similar improvement in $\dot{G}/G$ .

\section*{Acknowledgements}

The MeerKAT telescope is operated by the South African Radio Astronomy Observatory, which is a facility of the National Research Foundation, an agency of the Department of Science and Innovation.  SARAO acknowledges the ongoing advice and calibration of GPS systems by the National Metrology Institute of South Africa (NMISA) and the time space reference systems department of the Paris Observatory. \texttt{PTUSE} was developed with support from the Australian SKA Office and Swinburne University of Technology, with financial contributions from the MeerTime collaboration members. This work used the OzSTAR national facility at Swinburne University of Technology. OzSTAR is funded by Swinburne University of Technology and the National Collaborative Research Infrastructure Strategy (NCRIS). The National Radio Astronomy Observatory is a facility of the National Science Foundation operated under cooperative agreement by Associated Universities, Inc. V.V.K acknowledges financial support from the European Research Council (ERC) starting grant "COMPACT" (Grant agreement number 101078094). Parts of this research were conducted by the Australian Research Council Centre of Excellence for Gravitational Wave Discovery (OzGrav), through project number CE170100004.




\bibliographystyle{aa}
\bibliography{main_paper} 






\end{document}